\documentclass{pasa}
\usepackage{longtable}
\usepackage{hyperref}

\title[Long-term optical $BVRI$ variability of V2492 Cygni]{V2492 Cygni: Optical $BVRI$ variability during the period 2010$-$2017}
\author[Ibryamov et al.]{Sunay I. Ibryamov$^{1}$\thanks{E-mail: sibryamov@shu.bg}, Evgeni H. Semkov$^2$, Stoyanka P. Peneva$^2$\\
\affil{$^1$Department of Physics and Astronomy, University of Shumen, 115, Universitetska Str., 9700 Shumen, Bulgaria}
							\affil{$^2$Institute of Astronomy and National Astronomical Observatory, Bulgarian Academy of Sciences,
              72, Tsarigradsko Shose Blvd., 1784 Sofia, Bulgaria}}%
\jid{PASA}
\doi{10.1017/pas.\the\year.xxx}
\jyear{\the\year}

\begin{document}

\begin{abstract}
Results from $BVRI$ photometric observations of the young stellar object V2492 Cyg collected during the period from August 2010 to December 2017 are presented.
The star is located in the field of the Pelican Nebula and it was discovered in 2010 due to its remarkable increase in the brightness by more than 5 mag in $R$-band.
According to the first hypothesis of the variability V2492 Cyg is an FUor candidate. During subsequent observations it was reported that the star shows the characteristics inherent to EXor- and UXor-type variables.
The optical data show that during the whole time of observations the star exhibits multiple large amplitude increases and drops in the brightness.
In the beginning of 2017 we registered a significant increase in the optical brightness of V2492 Cyg, which seriously exceeds the maximal magnitudes registered after 2010.
\end{abstract}
\maketitle%
\section{INTRODUCTION}
\label{sec:intro}

During the Pre-main sequence (PMS) evolution the young stellar objects show various types of stellar activity.
PMS stars are separated into two main types: low-mass (M $\leq$ 2M$_{\odot}$) T Tauri stars (TTS) and the more massive (2M$_{\odot}$ $\leq$ M $\leq$ 8M$_{\odot}$) Herbig Ae/Be stars (HAEBES) (Joy 1945; Herbst et al. 1994; Hillenbrand et al. 1992; Petrov 2003).
A special kind of variability can be seen in the PMS stars, which undergo outbursts and/or obscurations.
The outbursts of the PMS stars - with amplitude reaching up to 5 mag - are grouped into two types: FUors with the prototype FU Orionis (Reipurth \& Aspin 2010; Audard et al. 2014) and EXors with the prototype EX Lupi (Herbig 2007; Reipurth \& Aspin 2010).
Such outbursts are rare occurrences and they are related to PMS stars during their evolutionary state, which ranges from early deeply embedded Class 0/I stage to a late Class II-type object.
Theoretical models (see Vorobyob \& Basu 2015) predict that after a sequence of eruptions the stars enter a more quiescent phase becoming classical T Tauri stars.
Also, according to Feh\'{e}r et al. (2017), FUors might represent the link between Class I and Class II low-mass young stars and the FUor outbursts may play the key role in this transition.

The outbursts of both types of eruptive PMS stars are generally attributed to a sizable increase in the accretion rate from the circumstellar disk onto the stellar surface.
In the case of FUors the accretion rate rapidly increases and remains elevated over several decades or more.
EXors exhibit shorter and repetitive outbursts associated with lower accretion rates (Audart et el. 2014).
Spectroscopic observations of EX Lupi itself in outbursts and during its quiescence phases revealed that it is an M0 classical T Tauri star that suffers episodes of variable mass accretion (Herbig et al. 2001, Herbig 2007).

The drops in the brightness of the PMS stars with amplitude reaching up to 3 mag in $V$-band are mostly seen in the early types of TTS and in HAEBES (Grinin et al. 1991, Herbst et al. 2007).
Such stars are called UXors, named after their prototype UX Orionis.
Different members of this group show different photometric activity which can change over time (Zaitseva 1986).
It is generally accepted that the observed minima result from the variations in the density of the dust in the orbit around the star, which crosses the line of sight and obscures the star.
This idea was first proposed by Wenzel (1969) and then discussed in further research by Grinin (1988), Voshchinnikov (1989), Grinin et al. (1991), Natta \& Whitney (2000), Dullemond et al. (2003). 

An important characteristic of UXors is that in their deep minima these objects become bluer - the so-called ``blueing effect'' or ``colour reverse'' (Bibo \& Th\'{e} 1990).
The interpretation of this effect states that the star is surrounded by circumstellar clouds and/or cometary bodies (Grady et al. 2000).
When one of these objects crosses the line of sight, a decrease in star's brightness is observed. 
Because of the absorption, the star initially becomes redder, but in a large extinction the scattered light from the dust clouds begins to dominate and the star becomes bluer.
The interferometric millimeter observations of some UXors and their analysis show that these stars are surrounded by circumstellar disks similar to those around TTS: optically thick with a mass 0.01$-$0.1 M$_{\odot}$ (Natta et al. 1999).
A large number of UXors are found to be HAEBES, but among the low-mass objects also there are representatives exhibiting the characteristics of UXors, for example see V582 Aur (Semkov et al. 2013) and V350 Cep (Semkov et al. 2017).

Variability of V2492 Cyg (also known as IRAS 20496+4354, PTF 10nvg and VSX J205126.1+440523) was discovered by Itagaki \& Yamaoka (2010).
On the basis of the photometric and spectral observations of the object, Covey et al. (2011) concluded that its brightening is indicative of enhanced accretion and outflow similar to the behavior of V1647 Ori in 2004-2005.
According to Aspin (2011) the spectral characteristics of V2492 Cyg during its outburst in 2010 are very similar to those exhibited by EX Lupi during its 2008 outburst, and V1647 Ori during its elevated phase in 2013.
Their conclusion is that V2492 Cyg is similar to EX Lupi although apparently significantly younger.
K\'{o}sp\'{a}l et al. (2013) concluded that a single physical mechanism is responsible for the brightness changes of the object and the colour variations suggest that the most likely explanation is the changing extinction along the line of sight.
Hillenbrand et al. (2013) discussed the object as a possible source exhibiting both accretion- and extinction-driven high-amplitude variability phenomena.
Aspin (2011) reported periodicity of $\sim$100 d for V2492 Cyg, while Hillenbrand et al. (2013) found $\sim$220 d quasi-periodic signal from the object.

In this paper we present our optical $BVRI$ observations of V2492 Cyg and discuss its light curves and colour-magnitude diagrams.
Using ours and the available archival observations and data in the literature and in the AAVSO database we built the long-term multicolour photometric light curves of the object.

\section{OBSERVATIONS AND CALIBRATION OF THE STANDARD STARS}

The photometric $BVRI$ observations of V2492 Cyg were carried out with the 2-m Ritchey-Chr\'{e}tien-Coud\'{e} (RCC), the 50/70-cm Schmidt and the 60-cm Cassegrain telescopes administered by Rozhen National Astronomical Observatory in Bulgaria and the 1.3-m Ritchey-Chr\'{e}tien (RC) telescope administered by Skinakas Observatory\footnote{Skinakas Observatory is a collaborative project of the University of Crete, the Foundation for Research and Technology, Greece, and the Max-Planck-Institut f{\"u}r Extraterrestrische Physik, Germany.} in Greece.

Our CCD observations were performed from August 2010 to December 2017.
The technical parameters and specifications for the cameras used, the observational procedure and the data reduction process are given in Ibryamov et al. (2015).
All data were analyzed using the same aperture, which was chosen to have a 4 arcsec radius, while the background annulus was taken from 9 arcsec to 14 arcsec. 

A sequence of six comparison stars labeled from A to F in the field of V2492 Cyg was calibrated in the $VRI$-bands by K\'{o}sp\'{a}l et al. (2011). 
Since these stars are relatively bright and there is no $B$-band - we decided to calibrate new $BVRI$ sequence of comparison stars including stars from B to F of K\'{o}sp\'{a}l et al. (2011).
We chose six more single stars in the field of V2492 Cyg and calibrated all of them in the $BVRI$-bands of the standard Johnson-Cousins system.
Calibration was made during eight clear nights (26 August 2010, 17 August 2011, 10 September 2011, 19 September 2011, 03 September 2012, 09 September 2012, 22 September 2012 and 12 August 2015) with the 1.3-m RC telescope.
Standard stars from Landolt (1992) were used as a reference.
Table 1 contains the photometric data for the $BVRI$ comparison sequence.
The corresponding average errors in the table are also listed.
The stars are labeled from B to L in order of their $V$-band magnitude and retaining the marks from B to F for the stars from the paper of K\'{o}sp\'{a}l et al. (2011). 
By comparing our measured magnitudes with those of K\'{o}sp\'{a}l et al. (2011) we get a very good match for $V$- and relatively significant differences (0.1-0.2 mag) for $R$- and $I$-band.
Due to the relative major errors from measurements of the star F we suspect that it is a possible variable with small amplitude and we advise the observers to use it with discretion.

The chart with the findings of the comparison sequence is presented in Fig. 1.
The magnitudes of V2492 Cyg from all CCD frames are measured with the standard stars reported in the present paper.
The average value of the errors in the reported magnitudes are 0.01$-$0.02 mag for the $I$- and $R$-band data, 0.01$-$0.03 mag for the $V$-band data and 0.02$-$0.04 mag for the $B$-band data.

\begin{table*}
\caption{Photometric data for the $BVRI$ comparison sequence.} 
\begin{center}
\begin{tabular*}{\textwidth}{@{}c\x c\x c\x c\x c\x c\x c\x c\x c\x @{}}
\hline \hline
Star & $I$ & $\sigma_{I}$ & $R$ & $\sigma_{R}$ & $V$ & $\sigma_{V}$ & $B$ & $\sigma_{B}$ \\
\hline 
B & 10.741 & 0.049 & 12.538 & 0.020 & 14.242 & 0.021 & 16.877 & 0.025 \\
C & 13.092 & 0.031 & 13.925 & 0.016 & 14.683 & 0.016 & 15.907 & 0.022 \\
D & 13.218 & 0.028 & 14.178 & 0.017 & 14.968 & 0.014 & 16.237 & 0.015 \\
E & 14.161 & 0.028 & 14.733 & 0.016 & 15.289 & 0.014 & 16.241 & 0.018 \\
F & 13.634 & 0.053 & 14.792 & 0.078 & 15.860 & 0.091 & 17.553 & 0.101 \\
G & 15.100 & 0.037 & 15.922 & 0.018 & 16.809 & 0.019 & 18.209 & 0.031 \\
H & 14.646 & 0.031 & 15.937 & 0.021 & 17.085 & 0.027 & 18.915 & 0.043 \\
I & 14.521 & 0.035 & 16.028 & 0.020 & 17.305 & 0.033 & 19.219 & 0.032 \\
J & 15.329 & 0.033 & 16.519 & 0.018 & 17.510 & 0.021 & 19.100 & 0.067 \\
K & 14.536 & 0.038 & 16.243 & 0.022 & 17.847 & 0.056 & 20.249 & 0.080 \\
L & 15.660 & 0.033 & 17.027 & 0.020 & 18.229 & 0.037 & 20.108 & 0.095 \\
\hline \hline
\end{tabular*}\label{Tab1}
\end{center}
\end{table*}

\begin{figure}
\begin{center}
\includegraphics[width=\columnwidth]{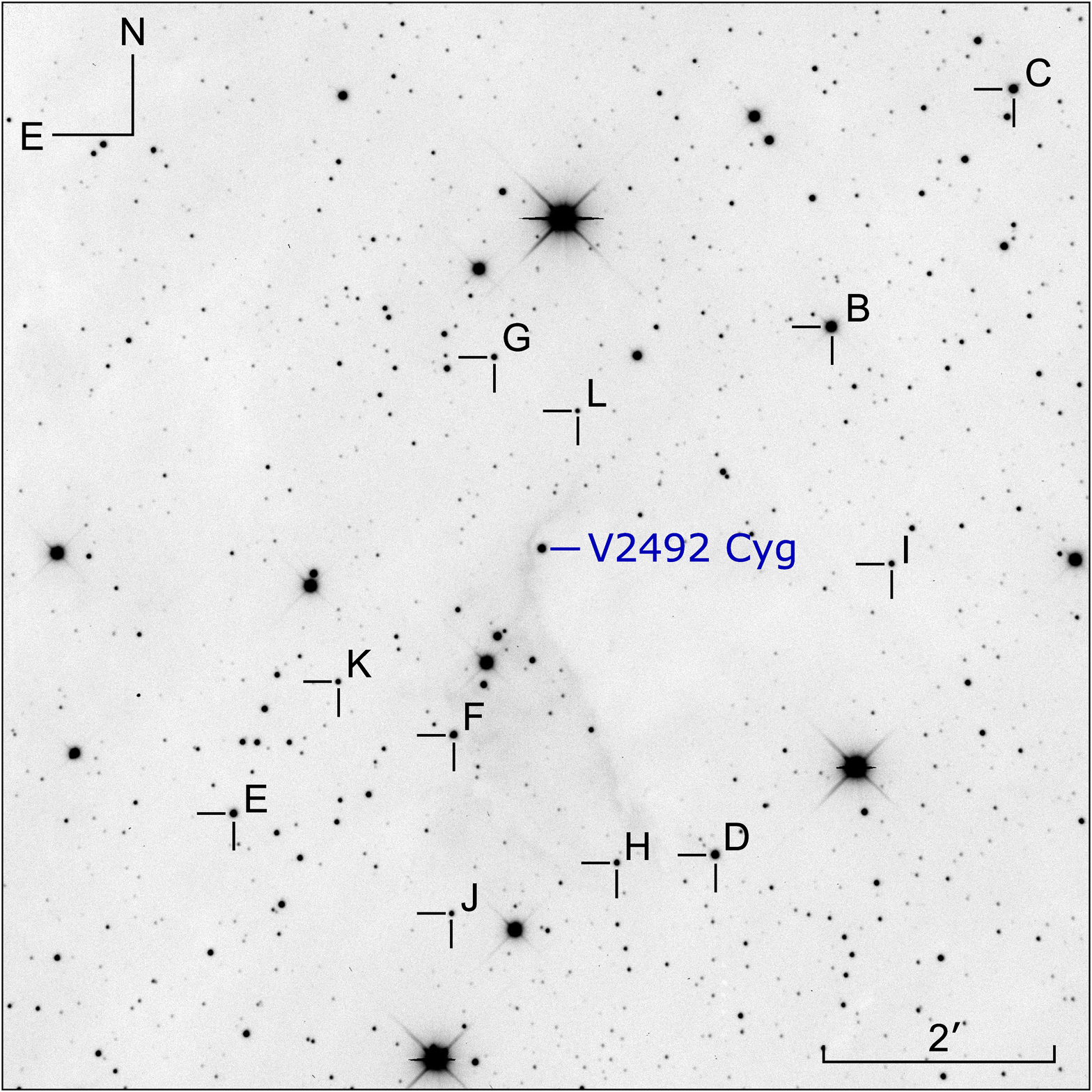}
\caption{Finding chart for the $BVRI$ comparison sequence around V2492 Cyg received in $R$-band with the 1.3-m RC telescope.}
\label{Fig1}
\end{center}
\end{figure}

\section{RESULTS AND DISCUSSION}

The optical photometric $BVRI$ data of V2492 Cyg received during our photometric monitoring and the ones available in the literature and in the AAVSO database are presented in Fig. 2.
The results of our CCD observations are summarized in Table A1\footnote{The table is available also via CDS VizieR Online Data Catalog.}. 

\begin{figure*}
\begin{center}
\includegraphics[width=\textwidth]{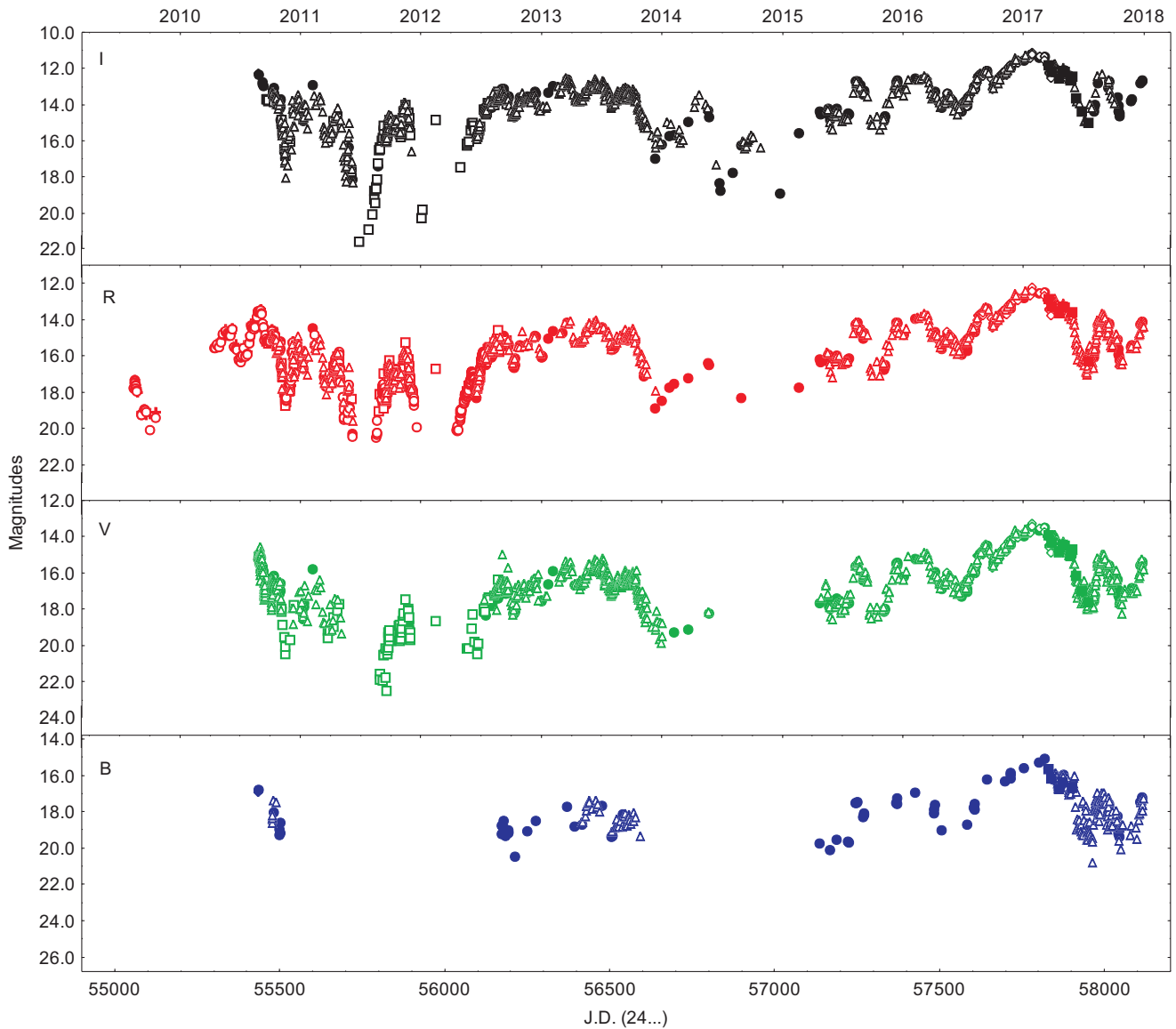}
\caption{$BVRI$ light curves of V2492 Cyg for the period August 2010 $-$ December 2017. The circles denote our CCD photometric data (present paper); the empty triangles represent the data from AAVSO database (\href{http://www.aavso.org}{http://www.aavso.org}); the diamonds represent the photometric data from Munari et al. (2010, 2017); the pluses denote the data from Covey et al. (2011); the empty squares mark the data from K\'{o}sp\'{a}l et al. (2011, 2013); the empty circles represent the data from Hillenbrand et al. (2013); the empty diamonds signify the data from Froebrich et al. (2017); and the squares represent the data from Giannini et al. (2017).}
\label{Fig2}
\end{center}
\end{figure*}

It can be seen from Fig. 2 that during the period 2010$-$2017 the object shows strong optical variability: in its light curves multiple increases and fades are visible, which are with different amplitudes and duration.
Several clearly expressed large amplitude increases and deep minima in the star's photometric behavior can be distinguished.
Between these events the star also shows short increases and drops in the brightness with smaller amplitudes.
It should be borne in mind, however, that in a very low light the star's brightness is under the photometric limit of the used telescopes. 
After February 2015 a deep minimum in the star's optical light curves is not observed.

Since the beginning of 2015 the brightness of V2492 Cyg began to rise, continuing to the beginning of 2017 when we registered a significant increase in its optical brightness, which seriously exceeds the maximal magnitudes registered after 2010 (reported in Ibryamov \& Semkov 2017).
The measured maximal values of the star's brightness were $B$=15.11 mag, $V$=13.52 mag, $R$=12.51 mag and $I$=11.37 mag.
After our report, Munari et al. (2017) obtained $BVRI$ photometry of V2492 Cyg and reported $B$=15.68 mag, $V$=13.93 mag, $R$=12.88 and $I$=11.84 mag, i.e. the object passed the recent maximum.
Giannini et al. (2017) presented optical and NIR photometric and spectroscopic observations of the star, which were obtained during the peak luminosity in 2016-2017.
Their combined analysis suggests that the V2492 Cyg variability is a combination of changing extinction and accretion.

An important result of our study is the change in the colour of V2492 Cyg.
Using data from our $BVRI$ photometry and the ones available in the literature three colour-magnitude diagrams ($B-V$/$V$, $V-R$/$V$ and $V-I$/$V$) of the object are constructed and displayed in Fig. 3.
It can be seen that the star becomes redder when fainter in a manner consistent with the extinction.
In accordance with the model of dust clumps obscuration the observed colour of the star is produced by the scattered light from the small dust grains.
Normally the star becomes redder when its light is covered by dust clumps or filaments in the line of sight.
Despite the fact that according to K\'{o}sp\'{a}l et al. (2013) the geometry of the system is closer to edge-on than to pole-on, and the observed fades events with large amplitudes, the blueing effect is not seen in the figure.
Most probably the obscuration does not rise sufficiently for the scattered light to become considerable enough for us to observe the star to become bluer.
In this case the variability of V2492 Cyg is likely dominated by variable accretion, but that does not rule out the obscuration by dust clumps or filaments in the environment of the object, proof of which is the dependence colour/magnitude shown in Fig. 3.

\begin{figure*}
\begin{center}
\includegraphics[width=4.0cm]{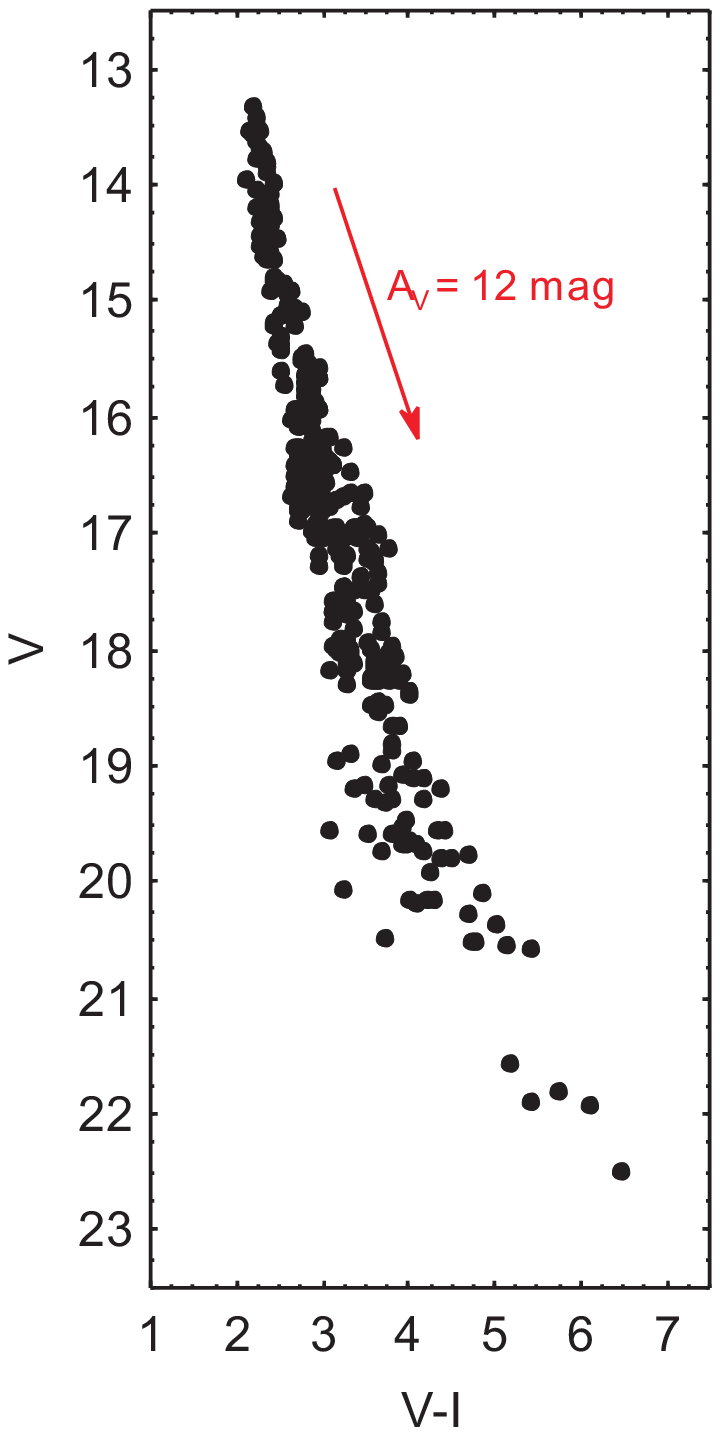}
\includegraphics[width=4.0cm]{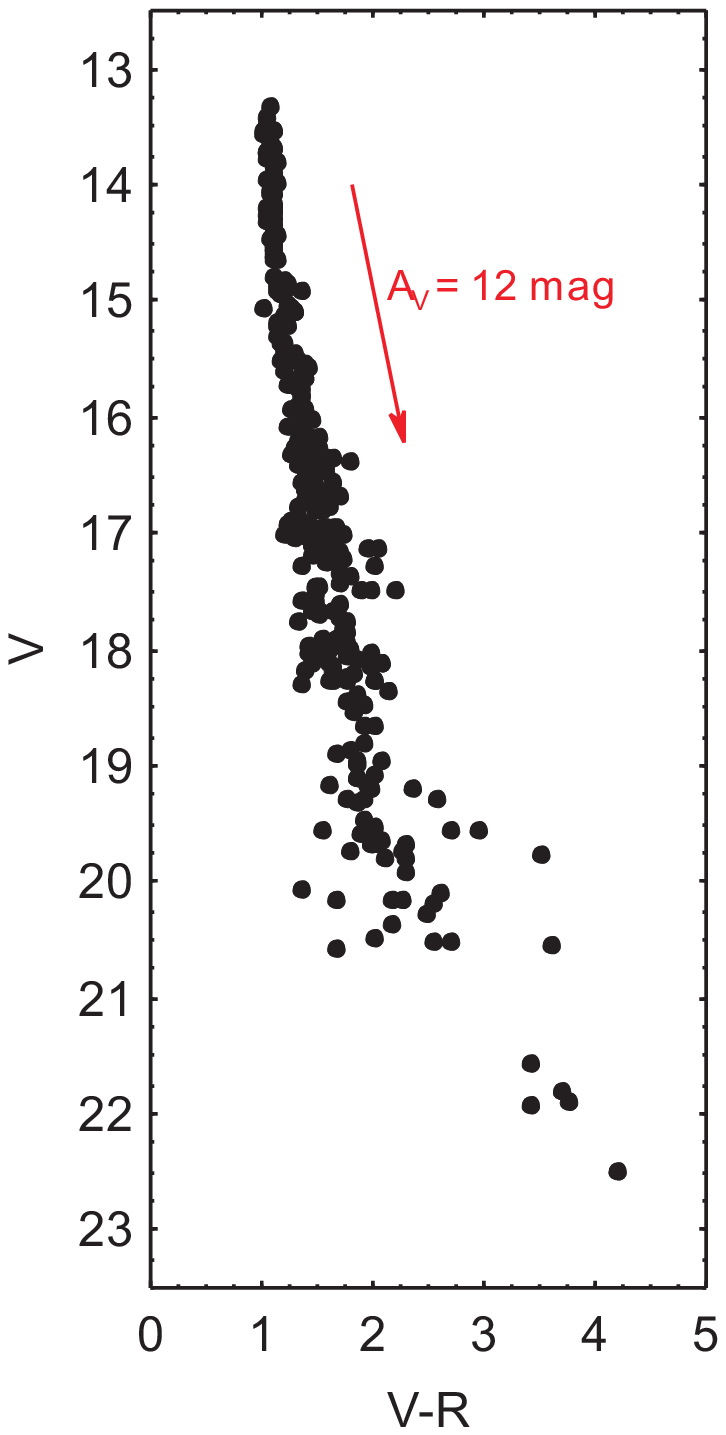}
\includegraphics[width=4.0cm]{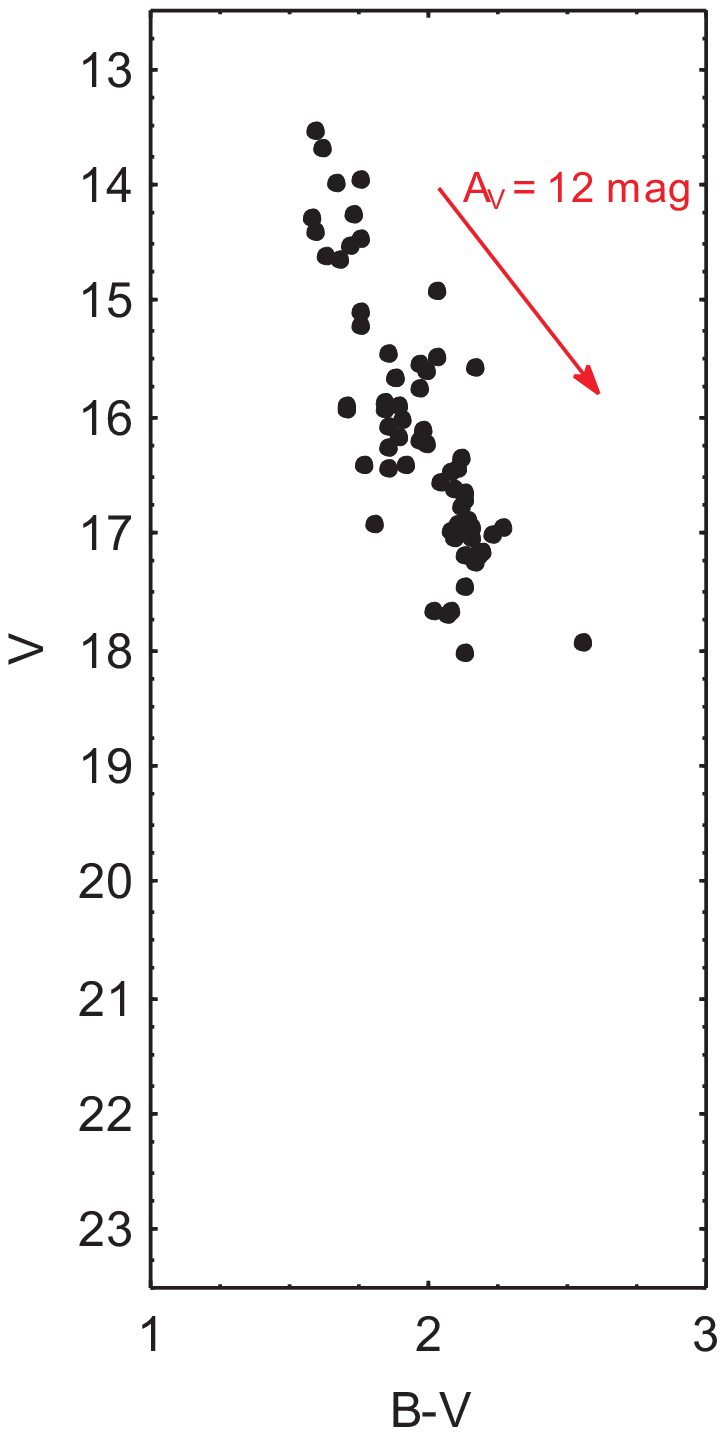}
\caption{Colour indices $V-I$, $V-R$ and $B-V$ versus the stellar $V$ magnitude of V2492 Cyg. The red line shows the reddening path for $A_{V}$=12 mag according to interstellar extinction law from Schlafly \& Finkbeiner (2011).}\label{Fig5}
\end{center}
\end{figure*}

The large amplitude outburst peak of V2492 Cyg in 2010, reported by Itagaki \& Yamaoka (2010), can only be explained as an episode of enhanced accretion.
The registered historical rise in the brightness of the object in the beginning of 2017 may be due to an increase in the accretion rate.
Another possible cause is the significant decrease of the extinction from clumps of dust orbiting the star.

We used the software packages \textsc{period04} (Lenz \& Breger 2005) and \textsc{persea} version 2.6 (written by G. Maciejewski on the ANOVA technique; Schwarzenberg-Czerny 1996) to search for periodicity in the light curves of V2492 Cyg.
Initially, we used all data points to search for periodicity in the light curves of the object but we did not find any periodicity.
After that we used data points from different periods of observations for the time-series analysis.
Only using the photometric data, received during the period from July 2012 to November 2013, our time-series analysis indicates a 99 day period and led to the ephemeris:

\begin{equation}\label{eq1}
JD (Max) = 2456374.998 + 99.099 * E.
\end{equation}

$I$-band folded light curve of V2492 Cyg according to the ephemeris (1) is plotted on Fig. 4.
False Alarm Probability (FAP) estimation was done by randomly deleting about 15$\%$ of the data about 50 times and then redetermining the period.
The period and starting age determinations remained stable even when a sub-sample of about 20$\%$ of the data were removed.
The obtained value for FAP is 0.02.
The result of our search for periodicity confirms the period of $\sim$100 d found by Aspin (2011).
The reason for not being able to find periodicity using all photometric data is probably hidden in the irrelevant mechanisms (accretion and obscuration) causing changes in the star's brightness in the different time-periods; it is also possible the periodicity of the object changes over time.

\begin{figure}
\begin{center}
\includegraphics[width=\columnwidth]{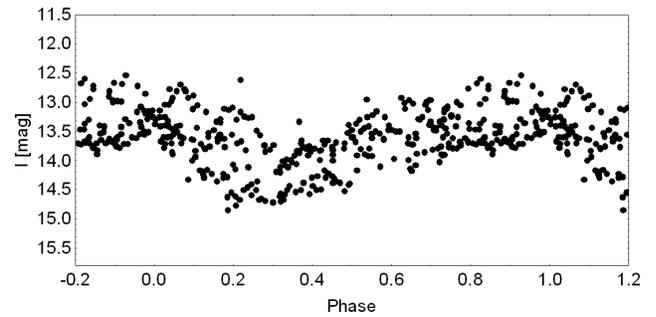}
\caption{$I$-band folded light curve of V2492 Cyg.}
\label{Fig4}
\end{center}
\end{figure}

After analysis of the collected data our conclusion is that the photometric properties of V2492 Cyg can be explained by superposition of both: (a) variable accretion from the circumstellar disk onto the stellar surface and (b) obscuration from circumstellar clumps of dust.
Both types of variability can act independently during different time periods and the result is the complicated light curve of V2492 Cyg.
A similar superposition of the both types of variability (accretion and obscuration) is seen on the long-term light curve of other PMS stars $-$ V582 Aurigae (Semkov et al. 2013), V1647 Orionis (Aspin et al. 2009), GM Cephei (Semkov et al. 2015a), and V1184 Tauri (Barsunova et al. 2006, Semkov et al. 2015b).

\section{CONCLUSION}
We built and investigated the long-term $BVRI$ photometric light curves of the young stellar object V2492 Cyg during the period from August 2010 to December 2017.
The photometric behavior of the object, including extreme large amplitudes and rapid variability, can be likely explained by a superposition of both accretion and obscuration events, although their mutual proportion has not been clarified yet.
We confirmed the $\sim$100 d periodicity in the light curves of the object, but it is available only using the photometric data received during the period from July 2012 to November 2013.

For determining the exact nature of V2492 Cyg further multicolour photometric and spectral observations are of great importance.
We plan to continue our photometric monitoring of the object during the next years.

\begin{acknowledgements}
This research has made use of the NASA's Astrophysics Data System.
The authors thank the Director of Skinakas Observatory Prof. I. Papamastorakis and Prof. I. Papadakis for the award of telescope time.
We acknowledge with thanks the variable star observations from the AAVSO International Database contributed by observers worldwide used in this research.
This work was partly supported by the Bulgarian Scientific Research Fund of the Ministry of Education and Science under the grants DM 08-2/2016, DN 08-1/2016, DN 08-20/2016 and DN 18-13/2017 as well as by the Scientific Research Fund of the University of Shumen.
We thank the anonymous referee for carefully reading the text and for the useful suggestions and comments that helped to improve the paper.
\end{acknowledgements}

\begin{appendix}

\section{$BVRI$-BANDS CCD PHOTOMETRY OF V2492 CYGNI}

\onecolumn
{\small
\begin{longtable}{cccccccc}
\caption{$BVRI$-bands CCD photometry of V2492 Cyg.}\\
\hline\hline
\noalign{\smallskip}  
Date (DD.MM.YYYY) & J.D. (24...) & I [mag] & R [mag] & V [mag] & B [mag] & Tel & CCD\\
\noalign{\smallskip}  
\hline
\endfirsthead
\caption{Continued.}\\
\hline\hline
\noalign{\smallskip}  
Date (DD.MM.YYYY) & J.D. (24...) & I [mag] & R [mag] & V [mag] & B [mag] & Tel & CCD\\
\noalign{\smallskip}  
\hline
\noalign{\smallskip}  
\endhead
\hline
\label{Tab2}
\endfoot
\noalign{\smallskip}
26.08.2010	&	55435.339	&	12.35	&	13.78	&	15.07	&	16.82	&	1.3-m	&	AND	\\
07.09.2010	&	55447.457	&	12.77	&	14.30	&	15.64	&	-	&	Sch	&	FLI	\\
08.09.2010	&	55448.298	&	12.87	&	14.40	&	15.74	&	-	&	Sch	&	FLI	\\
09.09.2010	&	55449.378	&	12.98	&	14.54	&	15.91	&	-	&	Sch	&	FLI	\\
11.10.2010	&	55481.316	&	13.09	&	14.66	&	16.17	&	18.06	&	1.3-m	&	AND	\\
29.10.2010	&	55499.210	&	14.22	&	15.73	&	17.17	&	19.30	&	2-m	&	VA	\\
30.10.2010	&	55500.184	&	14.00	&	15.49	&	16.89	&	18.99	&	2-m	&	VA	\\
31.10.2010	&	55501.178	&	13.71	&	15.27	&	-	&	-	&	Sch	&	FLI	\\
31.10.2010	&	55501.253	&	13.73	&	15.23	&	16.56	&	18.60	&	2-m	&	VA	\\
01.11.2010	&	55502.178	&	14.11	&	15.61	&	17.01	&	19.16	&	2-m	&	VA	\\
02.11.2010	&	55503.188	&	14.85	&	16.57	&	-	&	-	&	Sch	&	FLI	\\
03.11.2010	&	55504.219	&	14.97	&	16.76	&	-	&	-	&	Sch	&	FLI	\\
04.11.2010	&	55505.175	&	14.94	&	16.66	&	-	&	-	&	Sch	&	FLI	\\
05.11.2010	&	55506.206	&	14.82	&	16.52	&	17.94	&	-	&	Sch	&	FLI	\\
06.11.2010	&	55507.207	&	15.34	&	17.10	&	-	&	-	&	Sch	&	FLI	\\
01.01.2011	&	55563.188	&	14.17	&	16.12	&	-	&	-	&	Sch	&	FLI	\\
06.01.2011	&	55568.169	&	14.66	&	16.47	&	18.21	&	-	&	2-m	&	VA	\\
08.01.2011	&	55570.171	&	14.91	&	16.72	&	18.53	&	-	&	2-m	&	VA	\\
09.01.2011	&	55571.173	&	14.82	&	16.70	&	18.44	&	-	&	2-m	&	VA	\\
11.01.2011	&	55573.215	&	14.70	&	16.22	&	17.88	&	-	&	2-m	&	VA	\\
12.01.2011	&	55574.240	&	14.32	&	16.01	&	17.65	&	-	&	2-m	&	VA	\\
07.02.2011	&	55599.675	&	12.94	&	14.48	&	15.81	&	-	&	Sch	&	FLI	\\
04.04.2011	&	55656.484	&	15.50	&	-	&	-	&	-	&	Sch	&	FLI	\\
08.04.2011	&	55659.502	&	15.63	&	17.53	&	-	&	-	&	2-m	&	VA	\\
23.05.2011	&	55705.398	&	17.08	&	-	&	-	&	-	&	Sch	&	FLI	\\
25.05.2011	&	55707.379	&	16.39	&	-	&	-	&	-	&	Sch	&	FLI	\\
08.06.2011	&	55721.357	&	18.12	&	20.30	&	-	&	-	&	2-m	&	VA	\\
16.08.2011	&	55790.302	&	18.38	&	-	&	-	&	-	&	1.3-m	&	AND	\\
17.08.2011	&	55791.306	&	18.49	&	-	&	-	&	-	&	1.3-m	&	AND	\\
23.08.2011	&	55797.310	&	17.40	&	-	&	-	&	-	&	Sch	&	FLI	\\
10.09.2011	&	55815.428	&	15.34	&	18.11	&	-	&	-	&	1.3-m	&	AND	\\
11.09.2011	&	55816.415	&	15.38	&	18.20	&	20.36	&	-	&	1.3-m	&	AND	\\
19.09.2011	&	55824.258	&	15.92	&	18.67	&	-	&	-	&	1.3-m	&	AND	\\
23.09.2011	&	55828.268	&	15.25	&	-	&	-	&	-	&	Sch	&	FLI	\\
07.10.2011	&	55842.288	&	15.36	&	17.48	&	-	&	-	&	1.3-m	&	AND	\\
13.10.2011	&	55848.252	&	15.31	&	17.69	&	19.77	&	-	&	1.3-m	&	AND	\\
29.10.2011	&	55864.210	&	15.72	&	17.62	&	19.65	&	-	&	2-m	&	VA	\\
31.10.2011	&	55866.256	&	15.73	&	17.67	&	19.65	&	-	&	2-m	&	VA	\\
26.11.2011	&	55892.192	&	15.01	&	16.88	&	18.78	&	-	&	2-m	&	VA	\\
29.11.2011	&	55895.188	&	15.51	&	17.68	&	-	&	-	&	Sch	&	FLI	\\
30.11.2011	&	55896.201	&	15.63	&	17.89	&	-	&	-	&	Sch	&	FLI	\\
15.06.2012	&	56094.409	&	15.84	&	17.89	&	20.13	&	-	&	2-m	&	VA	\\
17.06.2012	&	56096.374	&	15.82	&	18.35	&	-	&	-	&	Sch	&	FLI	\\
11.07.2012	&	56120.350	&	14.17	&	16.24	&	17.94	&	-	&	Sch	&	FLI	\\
12.07.2012	&	56121.313	&	14.22	&	16.30	&	18.05	&	-	&	Sch	&	FLI	\\
13.07.2012	&	56122.375	&	14.27	&	16.38	&	18.18	&	-	&	Sch	&	FLI	\\
14.07.2012	&	56123.347	&	14.36	&	16.50	&	18.36	&	-	&	Sch	&	FLI	\\
01.08.2012	&	56141.399	&	13.99	&	15.90	&	17.59	&	-	&	1.3-m	&	AND	\\
02.08.2012	&	56142.285	&	14.18	&	16.10	&	17.84	&	-	&	1.3-m	&	AND	\\
03.08.2012	&	56143.269	&	14.09	&	15.98	&	17.74	&	-	&	1.3-m	&	AND	\\
12.08.2012	&	56151.614	&	13.71	&	15.54	&	17.22	&	-	&	1.3-m	&	AND	\\
19.08.2012	&	56159.326	&	13.68	&	15.59	&	17.25	&	-	&	Sch	&	FLI	\\
20.08.2012	&	56160.303	&	13.63	&	15.56	&	17.22	&	-	&	Sch	&	FLI	\\
21.08.2012	&	56160.540	&	13.61	&	15.48	&	17.20	&	-	&	1.3-m	&	AND	\\
21.08.2012	&	56161.323	&	13.78	&	15.74	&	17.42	&	-	&	Sch	&	FLI	\\
22.08.2012	&	56162.315	&	13.70	&	15.65	&	17.33	&	-	&	Sch	&	FLI	\\
02.09.2012	&	56173.310	&	13.35	&	15.26	&	16.99	&	19.21	&	1.3-m	&	AND	\\
03.09.2012	&	56174.274	&	13.15	&	15.01	&	16.63	&	18.75	&	1.3-m	&	AND	\\
07.09.2012	&	56178.280	&	13.15	&	14.90	&	16.46	&	18.53	&	1.3-m	&	AND	\\
08.09.2012	&	56179.449	&	13.26	&	15.06	&	16.67	&	-	&	1.3-m	&	AND	\\
09.09.2012	&	56180.281	&	13.33	&	15.14	&	16.75	&	18.86	&	1.3-m	&	AND	\\
10.09.2012	&	56181.267	&	13.46	&	15.29	&	16.94	&	19.08	&	1.3-m	&	AND	\\
11.09.2012	&	56182.222	&	13.43	&	15.29	&	16.93	&	19.08	&	1.3-m	&	AND	\\
12.09.2012	&	56183.346	&	13.59	&	15.47	&	17.15	&	19.34	&	1.3-m	&	AND	\\
22.09.2012	&	56193.254	&	13.56	&	15.36	&	16.95	&	19.03	&	1.3-m	&	AND	\\
22.09.2012	&	56193.317	&	13.56	&	15.36	&	16.92	&	19.18	&	Sch	&	FLI	\\
23.09.2012	&	56194.301	&	13.63	&	15.45	&	17.03	&	19.12	&	Sch	&	FLI	\\
07.10.2012	&	56208.229	&	14.62	&	16.62	&	18.26	&	-	&	Sch	&	FLI	\\
08.10.2012	&	56209.232	&	14.68	&	16.66	&	18.26	&	-	&	Sch	&	FLI	\\
09.10.2012	&	56210.214	&	14.51	&	16.48	&	18.09	&	-	&	Sch	&	FLI	\\
11.10.2012	&	56212.241	&	14.54	&	16.50	&	18.14	&	-	&	60-cm	&	FLI	\\
13.10.2012	&	56214.223	&	14.39	&	16.16	&	17.91	&	20.46	&	2-m	&	VA	\\
25.10.2012	&	56226.298	&	13.60	&	15.40	&	16.93	&	-	&	Sch	&	FLI	\\
26.10.2012	&	56227.406	&	13.68	&	15.46	&	17.00	&	-	&	Sch	&	FLI	\\
17.11.2012	&	56249.191	&	13.60	&	15.49	&	17.11	&	-	&	Sch	&	FLI	\\
18.11.2012	&	56250.201	&	13.54	&	15.38	&	16.95	&	19.08	&	Sch	&	FLI	\\
12.12.2012	&	56274.189	&	13.29	&	14.88	&	16.36	&	-	&	2-m	&	VA	\\
14.12.2012	&	56276.191	&	13.38	&	14.93	&	16.42	&	18.52	&	2-m	&	VA	\\
31.12.2012	&	56293.225	&	14.23	&	16.01	&	17.49	&	-	&	Sch	&	FLI	\\
01.01.2013	&	56294.205	&	14.28	&	16.09	&	17.56	&	-	&	60-cm	&	FLI	\\
03.01.2013	&	56296.253	&	14.15	&	16.02	&	-	&	-	&	60-cm	&	FLI	\\
19.01.2013	&	56312.197	&	13.34	&	15.05	&	16.63	&	-	&	2-m	&	VA	\\
05.02.2013	&	56329.182	&	12.96	&	14.61	&	15.91	&	-	&	Sch	&	FLI	\\
06.03.2013	&	56357.621	&	13.04	&	14.75	&	16.25	&	-	&	60-cm	&	FLI	\\
18.03.2013	&	56369.585	&	12.63	&	14.17	&	15.57	&	17.73	&	2-m	&	VA	\\
12.04.2013	&	56394.505	&	13.56	&	15.27	&	16.69	&	18.81	&	Sch	&	FLI	\\
04.05.2013	&	56417.438	&	13.66	&	15.22	&	16.60	&	18.69	&	2-m	&	VA	\\
15.05.2013	&	56428.423	&	13.26	&	14.81	&	16.24	&	-	&	60-cm	&	FLI	\\
17.05.2013	&	56430.426	&	-	&	14.63	&	-	&	-	&	60-cm	&	FLI	\\
30.05.2013	&	56443.403	&	13.00	&	14.56	&	15.91	&	17.75	&	Sch	&	FLI	\\
31.05.2013	&	56444.375	&	12.97	&	14.50	&	15.85	&	17.69	&	Sch	&	FLI	\\
04.07.2013	&	56478.389	&	12.98	&	14.42	&	15.74	&	17.70	&	2-m	&	VA	\\
01.08.2013	&	56506.376	&	14.02	&	15.68	&	17.23	&	19.39	&	2-m	&	VA	\\
02.08.2013	&	56507.386	&	13.98	&	15.62	&	17.18	&	19.36	&	2-m	&	VA	\\
03.08.2013	&	56508.395	&	13.99	&	15.63	&	17.17	&	19.33	&	2-m	&	VA	\\
04.08.2013	&	56509.310	&	13.92	&	15.63	&	17.17	&	-	&	Sch	&	FLI	\\
05.08.2013	&	56510.365	&	13.87	&	15.62	&	17.07	&	-	&	Sch	&	FLI	\\
05.08.2013	&	56510.392	&	13.84	&	15.56	&	17.01	&	-	&	60-cm	&	FLI	\\
06.08.2013	&	56511.432	&	13.77	&	15.49	&	17.01	&	-	&	60-cm	&	FLI	\\
07.08.2013	&	56512.379	&	13.81	&	15.57	&	17.03	&	-	&	Sch	&	FLI	\\
07.08.2013	&	56512.423	&	13.78	&	15.51	&	16.99	&	-	&	60-cm	&	FLI	\\
08.08.2013	&	56513.401	&	13.85	&	15.57	&	16.94	&	-	&	60-cm	&	FLI	\\
09.08.2013	&	56514.369	&	13.80	&	15.50	&	16.94	&	-	&	60-cm	&	FLI	\\
04.09.2013	&	56540.298	&	13.29	&	14.96	&	16.38	&	18.29	&	Sch	&	FLI	\\
05.09.2013	&	56541.292	&	13.26	&	14.95	&	16.38	&	18.14	&	Sch	&	FLI	\\
07.09.2013	&	56543.426	&	13.34	&	14.89	&	16.35	&	18.46	&	2-m	&	VA	\\
08.09.2013	&	56544.280	&	13.48	&	15.05	&	16.48	&	-	&	2-m	&	VA	\\
11.09.2013	&	56547.408	&	13.11	&	14.74	&	16.15	&	-	&	60-cm	&	FLI	\\
17.09.2013	&	56553.262	&	13.46	&	15.16	&	16.67	&	-	&	1.3-m	&	AND	\\
11.10.2013	&	56577.327	&	13.59	&	15.19	&	16.57	&	-	&	60-cm	&	FLI	\\
12.10.2013	&	56578.352	&	13.53	&	15.13	&	16.54	&	-	&	60-cm	&	FLI	\\
07.11.2013	&	56604.286	&	15.27	&	17.11	&	-	&	-	&	60-cm	&	FLI	\\
09.12.2013	&	56636.209	&	17.00	&	18.91	&	-	&	-	&	2-m	&	VA	\\
29.12.2013	&	56656.202	&	16.22	&	18.46	&	-	&	-	&	Sch	&	FLI	\\
23.01.2014	&	56681.208	&	15.74	&	17.74	&	-	&	-	&	Sch	&	FLI	\\
06.02.2014	&	56694.641	&	15.70	&	17.54	&	19.28	&	-	&	2-m	&	VA	\\
22.03.2014	&	56738.568	&	14.93	&	17.25	&	19.10	&	-	&	Sch	&	FLI	\\
21.05.2014	&	56799.450	&	14.34	&	16.40	&	-	&	-	&	Sch	&	FLI	\\
23.05.2014	&	56801.411	&	14.70	&	16.51	&	18.26	&	-	&	2-m	&	VA	\\
23.06.2014	&	56832.439	&	18.37	&	-	&	-	&	-	&	2-m	&	VA	\\
25.06.2014	&	56834.400	&	18.75	&	-	&	-	&	-	&	2-m	&	VA	\\
03.08.2014	&	56873.383	&	17.77	&	-	&	-	&	-	&	2-m	&	VA	\\
29.08.2014	&	56899.292	&	16.24	&	18.33	&	-	&	-	&	1.3-m	&	AND	\\
24.12.2014	&	57016.190	&	18.92	&	-	&	-	&	-	&	2-m	&	VA	\\
21.02.2015	&	57074.619	&	15.60	&	17.77	&	-	&	-	&	Sch	&	FLI	\\
23.04.2015	&	57136.465	&	14.40	&	16.19	&	17.68	&	19.74	&	Sch	&	FLI	\\
25.04.2015	&	57138.472	&	14.54	&	16.36	&	-	&	-	&	Sch	&	FLI	\\
18.05.2015	&	57161.391	&	14.22	&	15.96	&	17.45	&	-	&	Sch	&	FLI	\\
21.05.2015	&	57164.440	&	14.60	&	16.36	&	17.88	&	-	&	Sch	&	FLI	\\
24.05.2015	&	57167.386	&	14.83	&	16.46	&	18.00	&	20.13	&	2-m	&	VA	\\
13.06.2015	&	57187.356	&	14.21	&	15.97	&	17.43	&	19.55	&	2-m	&	VA	\\
16.07.2015	&	57220.341	&	14.64	&	16.43	&	17.73	&	-	&	Sch	&	FLI	\\
17.07.2015	&	57221.393	&	14.47	&	16.22	&	17.57	&	-	&	Sch	&	FLI	\\
19.07.2015	&	57223.366	&	14.55	&	16.19	&	17.64	&	19.65	&	2-m	&	VA	\\
20.07.2015	&	57224.394	&	14.52	&	16.13	&	17.64	&	19.71	&	2-m	&	VA	\\
11.08.2015	&	57246.372	&	12.71	&	14.26	&	15.64	&	17.52	&	1.3-m	&	AND	\\
12.08.2015	&	57247.299	&	12.70	&	14.28	&	15.66	&	17.54	&	1.3-m	&	AND	\\
17.08.2015	&	57252.305	&	12.72	&	14.15	&	15.52	&	17.48	&	2-m	&	VA	\\
24.08.2015	&	57259.466	&	13.11	&	14.69	&	-	&	-	&	Sch	&	FLI	\\
25.08.2015	&	57260.364	&	12.92	&	14.51	&	-	&	-	&	Sch	&	FLI	\\
03.09.2015	&	57269.350	&	13.44	&	15.03	&	16.43	&	18.28	&	Sch	&	FLI	\\
04.09.2015	&	57270.433	&	13.33	&	14.87	&	16.19	&	18.15	&	2-m	&	VA	\\
05.09.2015	&	57271.411	&	13.34	&	14.83	&	16.22	&	18.21	&	2-m	&	VA	\\
06.09.2015	&	57272.406	&	13.25	&	14.74	&	16.09	&	18.07	&	2-m	&	VA	\\
03.11.2015	&	57330.215	&	14.90	&	16.76	&	18.15	&	-	&	Sch	&	FLI	\\
04.11.2015	&	57331.201	&	14.83	&	16.64	&	18.09	&	-	&	Sch	&	FLI	\\
05.11.2015	&	57332.204	&	14.79	&	16.61	&	18.04	&	-	&	Sch	&	FLI	\\
06.11.2015	&	57333.202	&	14.72	&	16.60	&	18.02	&	-	&	Sch	&	FLI	\\
07.11.2015	&	57334.199	&	14.66	&	16.51	&	17.98	&	-	&	Sch	&	FLI	\\
12.12.2015	&	57369.186	&	12.74	&	14.19	&	15.48	&	17.50	&	2-m	&	VA	\\
13.12.2015	&	57370.163	&	12.75	&	14.21	&	15.60	&	17.59	&	2-m	&	VA	\\
14.12.2015	&	57371.177	&	12.77	&	14.24	&	15.50	&	-	&	2-m	&	VA	\\
15.12.2015	&	57372.193	&	12.67	&	14.16	&	15.44	&	17.29	&	Sch	&	FLI	\\
02.01.2016	&	57390.173	&	13.20	&	14.83	&	16.19	&	-	&	Sch	&	FLI	\\
07.02.2016	&	57426.188	&	12.55	&	13.98	&	15.20	&	16.95	&	Sch	&	FLI	\\
04.04.2016	&	57483.498	&	13.54	&	14.96	&	16.24	&	18.09	&	2-m	&	VA	\\
06.04.2016	&	57484.539	&	13.38	&	14.83	&	16.06	&	17.91	&	2-m	&	VA	\\
06.04.2016	&	57485.499	&	13.27	&	14.69	&	15.93	&	17.63	&	Sch	&	FLI	\\
27.04.2016	&	57506.420	&	14.17	&	15.62	&	16.88	&	19.02	&	Sch	&	FLI	\\
13.05.2016	&	57522.406	&	13.60	&	15.06	&	16.30	&	-	&	Sch	&	FLI	\\
14.05.2016	&	57523.408	&	13.37	&	14.80	&	16.08	&	-	&	Sch	&	FLI	\\
31.05.2016	&	57540.384	&	13.89	&	15.37	&	-	&	-	&	2-m	&	VA	\\
25.06.2016	&	57565.442	&	14.36	&	15.93	&	17.28	&	-	&	Sch	&	FLI	\\
11.07.2016	&	57581.363	&	13.89	&	15.53	&	16.85	&	-	&	Sch	&	FLI	\\
12.07.2016	&	57582.391	&	13.96	&	15.59	&	16.89	&	18.69	&	Sch	&	FLI	\\
13.07.2016	&	57583.377	&	14.05	&	15.73	&	17.02	&	-	&	Sch	&	FLI	\\
01.08.2016	&	57602.356	&	13.04	&	14.57	&	15.89	&	17.78	&	2-m	&	VA	\\
04.08.2016	&	57605.360	&	13.14	&	14.67	&	16.00	&	17.90	&	Sch	&	FLI	\\
05.08.2016	&	57606.353	&	13.07	&	14.58	&	15.89	&	17.59	&	Sch	&	FLI	\\
11.09.2016	&	57643.261	&	12.16	&	13.39	&	14.50	&	16.21	&	Sch	&	FLI	\\
02.10.2016	&	57664.230	&	-	&	14.32	&	15.49	&	-	&	Sch	&	FLI	\\
05.11.2016	&	57698.260	&	12.21	&	13.49	&	14.63	&	16.31	&	Sch	&	FLI	\\
21.11.2016	&	57714.222	&	11.89	&	13.21	&	14.25	&	15.97	&	2-m	&	VA	\\
22.11.2016	&	57715.204	&	11.87	&	13.16	&	14.27	&	15.85	&	2-m	&	VA	\\
23.11.2016	&	57716.215	&	11.97	&	13.38	&	14.44	&	16.19	&	2-m	&	VA	\\
02.01.2017	&	57756.202	&	11.54	&	12.83	&	13.96	&	15.62	&	Sch	&	FLI	\\
17.02.2017	&	57801.603	&	11.40	&	12.57	&	13.67	&	15.28	&	Sch	&	FLI	\\
05.03.2017	&	57817.549	&	11.37	&	12.51	&	13.52	&	15.11	&	Sch	&	FLI	\\
02.04.2017	&	57845.538	&	12.28	&	13.50	&	14.61	&	16.24	&	Sch	&	FLI	\\
03.04.2017	&	57846.567	&	12.09	&	13.29	&	14.38	&	15.97	&	Sch	&	FLI	\\
01.05.2017  & 57875.483 & 12.10 & 13.38 & 14.43 & 15.96 & 2-m & VA  \\
18.05.2017	& 57892.420	& 12.58	& 13.85	& 15.01	& 16.75	& Sch	& FLI \\
19.05.2017	& 57893.422	& 12.42	& 13.75	& 14.83	& 16.37	& 2-m	& VA  \\
01.08.2017	& 57967.382	& 14.37	& 16.11	& 17.55	& -     &	Sch	& FLI \\
02.08.2017	& 57968.322	& 14.18	& 15.94	& 17.43	& -     &	Sch	& FLI \\
03.08.2017	& 57969.336	& 14.03	& 15.76	& 17.26	& -     &	Sch	& FLI \\
12.08.2017	& 57978.495	& 12.79	& 14.31	& 15.61	& 17.51	& Sch	& FLI \\
14.09.2017	& 58011.268	& 13.07	& 14.68	& 16.08	& 18.08	& Sch	& FLI \\
15.09.2017	& 58012.286	& 12.69	& 14.23	& 15.58	& 17.60	& Sch	& FLI \\
16.09.2017	& 58013.269	& 12.77	& 14.34	& 15.71	& 17.69	& Sch	& FLI \\
12.10.2017	& 58039.223	& 13.59	& 15.11	& 16.39	& 18.27	& Sch	& FLI \\
14.10.2017	& 58041.398	& 14.07	& 15.61	& 16.95	& 18.94	& 2-m	& VA  \\
16.10.2017	& 58043.234	& 14.39	& 15.95	& 17.24	& 19.12	& Sch	& FLI \\
16.10.2017	& 58043.272	& 14.39	& 15.92	& 17.27	& 19.23	& 2-m	& VA  \\
17.10.2017	& 58044.295	& 14.48	& 16.09	& 17.39	& 19.41	& Sch	& FLI \\
18.10.2017	& 58045.385	& 14.62	& 16.23	& 17.70	& -     &	Sch	& FLI \\
22.11.2017	& 58080.245	& 13.81	& 15.53	& 16.99	& -     &	Sch	& FLI \\
23.11.2017	& 58081.257	& 13.71	& 15.40	& 16.85 & -     &	Sch &	FLI \\
21.12.2017	& 58109.266	& 12.79	& 14.29	& 15.59	& 17.49	& Sch	& FLI \\
25.12.2017	& 58113.189	& 12.68	& 14.12	& 15.36	& 17.21	& Sch	& FLI \\
26.12.2017	& 58114.223	& 12.72	& 14.19	& 15.50	& 17.32	& Sch	& FLI \\
\hline \hline
\end{longtable}}


\end{appendix}

\end{document}